# Dual channels of helicity cascade in turbulent flows

Zheng Yan[1,2], Xinliang Li[1,2], Changping Yu[1,2], Shiyi Chen[3,4]

[1]LHD, Institute of Mechanics, Chinese Academy of Sciences, Beijing 100190, China

[2]School of Engineering Science, University of Chinese Academy of Sciences, Beijing 100049, China

[3]Department of Mechanics and Aerospace Engineering, Southern University of Science and Technology, Shenzhen, Guangdong 518055, China

[4]State Key Laboratory of Turbulence and Complex Systems, Center for Applied Physics and Technology, College of Engineering, Peking University, Beijing 100871, China

Helicity, as one of only two inviscid invariants in three-dimensional turbulence, plays an important role in the generation and evolution of turbulence. From the traditional viewpoint, there exists only one channel of helicity cascade similar to that of kinetic energy cascade. Through theoretical analysis, we find that there are two channels in helicity cascade process. The first channel mainly originates from vortex twisting process, and the second channel mainly originates from vortex stretching process. By analysing the data of direct numerical simulations of typical turbulent flows, we find that these two channels behave differently. The ensemble averages of helicity flux in different channels are equal in homogeneous and isotropic turbulence, while they are different in other type of turbulent flows. The second channel is more intermittent and acts more like a scalar, especially on small scales. Besides, we find a novel mechanism of hindered even inverse energy cascade, which could be attributed to the second-channel helicity flux with large amplitude.

Helicity exists in many natural phenomena, such as hurricanes, tornadoes, and rotating "supercell" thunderstorms in the atmosphere, Langmuir circulation in the ocean, and $\alpha$-effect and $\omega$-effect in the magnetic field[1,2]. In the past few decades, there have been numerous theoretical and numerical conclusions indicating that helicity could reduce the aerodynamic drag, nonlinearity of Navier-Stokes equations (NSEs), and improve the mixing effectiveness of reactants[1,3]. Helicity, the integral of the scalar product of velocity and vorticity, is the second inviscid invariant of the three-dimensional(3D) NSEs, which indicate that helicity cascade exists in 3D turbulent flows. Recently, a new research has shown that helicity is a conservative quantity even in viscous flows[4,5]. Helicity is a topological variable, which measures the degree of the linkage of the vortex lines in the flow field[6], and consists of linking, twisting and writhing[4].

The classical Richardson-Kolmogorov-Onsager picture of 3D turbulence is based on the concept of energy cascade, which ignores the topology of vortices[7]. Theoretically, there are two possibilities describing the dynamical properties of helicity and energy cascades. One is simultaneous energy and helicity cascades toward smaller scales, and the other is a pure helicity cascade with no cascade of energy, leading to broken -5/3 power law solutions in the turbulent magnetohydrodynamical, convective and atmospheric flows[8,9]. While many studies revealed that through direct numerical simulations (DNS) and shell model there exists a transfer of energy and helicity to small scales simultaneously in turbulent flows at a high Reynolds number[8,10-13]. In the process of the joint cascade of energy and helicity in helical turbulence, helicity flux is more intermittent than energy flux[14]. For rotating helical turbulence, helicity flux dominates the direct energy cascade to small scales and the direct helicity cascade is highly intermittent[15,16]. However, the impact of helicity on the decaying rate of turbulent flows works only in rotating flows[17].

The role of helicity in the behaviors of turbulent dynamic systems has been a controversial issue in the past decades. Previous studies argued that the helicity cascade is carried along locally and linearly by the energy cascade and it acts like a passive scalar [18]. Another argument insists that the helicity cascade has a dramatic effect on the energy cascade. For instances, helicity can impede the forward energy cascade and even promote the inversion of energy transfer, which could be explained as the helical bottleneck effect [19-25].

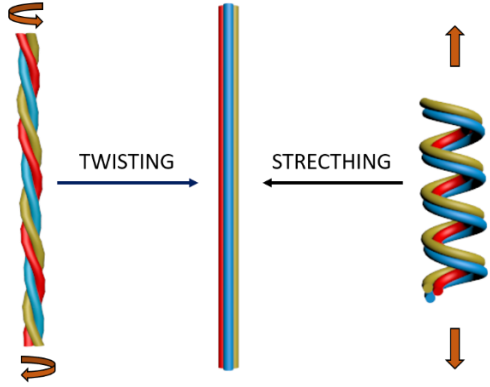

FIG.1. A schematic demonstrating two processes of vortex dynamics for helicity cascade, twisting and stretching.

Traditional theory reveals that there is only one channel of energy cascade in turbulent flows, and both forward and backward cascades exist in the same channel [26]. Moreover, the traditional view on helicity cascade is similar to that on energy cascade [11], and vortex twisting plays a major role in helicity cascade process [27]. In this letter, through theoretical and numerical investigations we discover that there exist two channels in the helicity cascade process, and they behave differently in turbulent flows.

In order to study the characteristics of the helicity cascade, we take the coarse-graining method to filter the flow field [28,29]. Using a smooth low-pass filter function $G_\Delta(r)$, we can obtain the filtered physical variable such as $\tilde{\mathbf{u}}(\mathbf{x}) = \int d\mathbf{r} G_\Delta(\mathbf{r}) \mathbf{u}(\mathbf{x}+\mathbf{r})$ representing the filtered velocity field on scale $\Delta$. The governing equations of large-scale energy $e_\Delta = |\tilde{\mathbf{u}}|^2/2$ and large-scale helicity $h_\Delta = \tilde{\mathbf{u}} \cdot \tilde{\boldsymbol{\omega}}$ could be easily obtained as follows,

$$\partial_t e_\Delta + \nabla \cdot \mathbf{J} = -\Pi_\Delta^E - 2\nu \tilde{\mathbf{S}} : \tilde{\mathbf{S}} + \tilde{\mathbf{f}} \cdot \tilde{\mathbf{u}} \tag{1}$$

$$\partial_t h_\Delta + \nabla \cdot \mathbf{Q} = -\Pi_\Delta^{H1} - \Pi_\Delta^{H2} - 4\nu \tilde{\mathbf{S}} : \tilde{\mathbf{R}} + \tilde{\mathbf{f}} \cdot \tilde{\boldsymbol{\omega}} + \tilde{\boldsymbol{\psi}} \cdot \tilde{\mathbf{u}} \tag{2}$$

Where $\tilde{\mathbf{u}}$ is the filtered velocity, $\tilde{\boldsymbol{\omega}}$ is the filtered vorticity, $\tilde{\mathbf{S}} = 1/2(\nabla \tilde{\mathbf{u}} + (\nabla \tilde{\mathbf{u}})^T)$, $\tilde{\boldsymbol{\Omega}} = 1/2(\nabla \tilde{\mathbf{u}} - (\nabla \tilde{\mathbf{u}})^T)$, $\tilde{\mathbf{R}} = 1/2(\nabla \tilde{\boldsymbol{\omega}} + (\nabla \tilde{\boldsymbol{\omega}})^T)$, $\tilde{\mathbf{f}}$ is the external force, $\tilde{\boldsymbol{\psi}} = \nabla \times \tilde{\mathbf{f}}$. Here the space transport of large-scale energy equation denotes $\mathbf{J} = \tilde{\mathbf{u}}|\tilde{\mathbf{u}}|^2/2 + \boldsymbol{\tau} \cdot \tilde{\mathbf{u}} - 2\tilde{\mathbf{S}} \cdot \tilde{\mathbf{u}}$, and the SGS

energy flux denotes $\Pi_\Delta^E = -\boldsymbol{\tau}:\tilde{\mathbf{S}}$. Here $\mathbf{Q}$ is the spatial transport of large-scale helicity, which is defined as

$$\mathbf{Q} = h_\Delta \tilde{\mathbf{u}} + \tilde{\boldsymbol{\omega}} \cdot \boldsymbol{\tau} + \tilde{\mathbf{u}} \cdot \boldsymbol{\gamma} + \frac{\tilde{p}}{\rho}\tilde{\boldsymbol{\omega}} - \frac{1}{2}|\tilde{\mathbf{u}}|^2 \tilde{\boldsymbol{\omega}} - 2\nu\left(\tilde{\mathbf{u}} \cdot \tilde{\mathbf{R}} + \tilde{\boldsymbol{\omega}} \cdot \tilde{\mathbf{S}}\right) \tag{3}$$

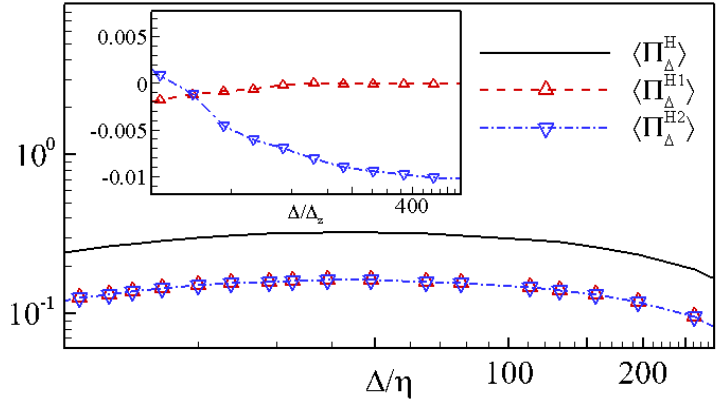

FIG.2. Ensemble averages of the first-channel, the second-channel and total helicity flux on the different filter width in HHIT. The first-channel and second-channel helicity flux on the plane of $y^+$ in turbulent channel flows are shown in the inset.

It needs to be mentioned specifically here that the first-channel helicity flux is named as $\Pi_\Delta^{H1}$ and the second-channel helicity flux is named as $\Pi_\Delta^{H2}$. They can be expressed as

$$\Pi_\Delta^{H1} = -\boldsymbol{\tau}:\tilde{\mathbf{R}}, \Pi_\Delta^{H2} = -\boldsymbol{\gamma}:\tilde{\boldsymbol{\Omega}} \tag{4}$$

where $\boldsymbol{\tau} = \mathbf{uu} - \tilde{\mathbf{u}}\tilde{\mathbf{u}}$ is the subgrid-scale (SGS) stress, and $\boldsymbol{\gamma} = (\boldsymbol{\omega}\mathbf{u} - \tilde{\boldsymbol{\omega}}\tilde{\mathbf{u}}) - (\boldsymbol{\omega}\mathbf{u} - \tilde{\boldsymbol{\omega}}\tilde{\mathbf{u}})^T$ can be called SGS vortex stretching stress. From the definition of the first and second channels, we can find that the first channel originates mainly from the vortex twisting process, and the second channel originates mainly from the vortex stretching process. Hence, we could conclude that the helicity cascade is a combined process of vortex twisting and stretching. These two processes are illustrated schematically in Fig.1.

As a further step to explore the statistical features of the dual-channel helicity cascade, we perform forced helical homogeneous and isotropic turbulence (HHIT) with $\text{Re}_\lambda = 341$ within a cubic box with sides of length $2\pi$ at a grid resolution of $1024^3$ by the pseudo spectral solver, and access DNS data of turbulent channel flows with $\text{Re}_\tau \approx 1000$ via the Johns Hopkins Turbulence Database [30]. In Fig.2, we present the dependence of the ensemble averages of the first-channel, second-channel and total helicity flux, respectively, on the different filter width. In HHIT, the ensemble averages of the first-channel and second-channel helicity flux are exactly equal in our numerical simulations. The equality of ensemble averages of these two channels in HHIT can be proved exactly by homogeneity according to the following identical relation and Gauss' flux theorem.

$$\nabla \cdot (\mathbf{a} \times \mathbf{b}) = \mathbf{b} \cdot (\nabla \times \mathbf{a}) - \mathbf{a} \cdot (\nabla \times \mathbf{b}) \tag{5}$$

where $\mathbf{a}$ and $\mathbf{b}$ are two arbitrary vectors.

However, the equality of these two-channel fluxes will be broken in other flows like turbulent channel flows with $\text{Re}_\tau \approx 1000$ on the plane $y^+ = 1$ [30] in the inset of Fig.2. In order to explore the features of spatial distribution further, the ensemble averages of the first- and second-channel helicity flux and energy flux depending on scales and distances from the wall are exhibited in Fig.3. It is shown that the local interactions along with transfer of energy and helicity are drastic in the buffer layer, and the distributions of the two channels of helicity cascade are evidently fluctuating relative to energy flux. We also find that the first channel dominates the helicity transfer, which can reveal that the vortex twisting effect plays an important role in the helicity cascade process.

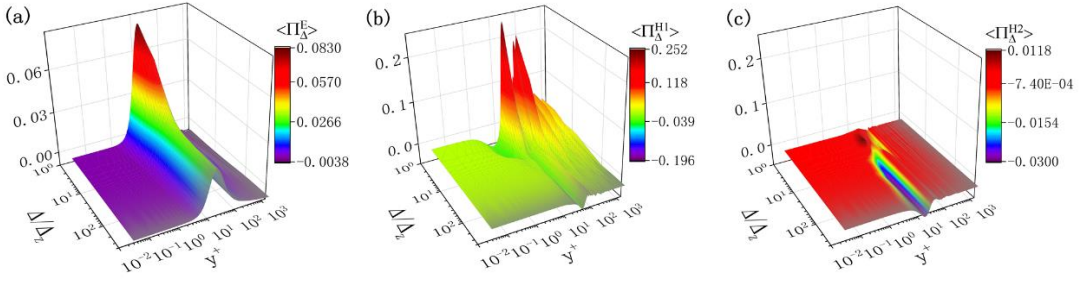

FIG.3. Ensemble averages of the first- and second-channel helicity flux and energy flux on different length scales ($\Delta/\Delta_z$) and distances from the wall ($y^+$).

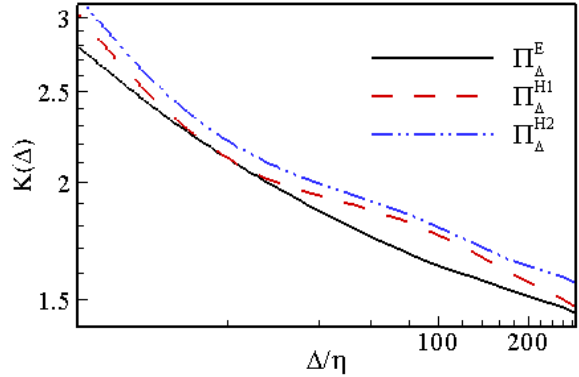

FIG.4. Excess kurtosis of the energy flux, first-channel and the second-channel helicity flux.

However, these two channels of helicity cascade in HHIT have different statistical properties in higher statistical order. Their normalized fourth orders are chosen to illustrate their statistical discrepancy, which is related to the intermittency representing the strong non-Gaussian fluctuations. It could be assessed quantitatively by excess kurtosis [14,31]. In Fig.4, we exhibit the excess kurtosis of the first- and second-channel helicity flux and energy flux, respectively. Apparently, the excess kurtosis of the second-channel helicity flux is larger than that of the first-channel helicity flux, and the excess kurtosis of first-channel helicity flux is larger than that of energy flux. It means that the first-channel helicity flux is more intermittent than energy flux, which is consistent with the conclusion in reference [14]. Naturally, we see that the second-channel helicity flux is also more intermittent than the first-channel helicity flux. In order to explore the discrepancy of intermittency in detail, we show their normalized probability density functions (PDF) at filter width $\Delta=24\eta$ in Fig.5. Previous studies revealed that a nearly symmetric distribution of helicity flux exists [14], and then we find that the distribution of the second-channel helicity flux is more symmetric than that of the first-channel helicity flux from Fig.5. The distribution regularities are same in the first and second channel projected on left and right chirality through helical wave decomposition, which are not shown for the sake of simplicity. The phenomenology of small-scale turbulence reveals that the scalar is more intermittent than advecting velocity [32], and thus we can see that the second channel acts more like a scalar.

As a further step to compare their spatial distribution and morphological characteristics in HHIT, isosurfaces of the energy flux ($\Pi_\Delta^E$) and the first- and second-channel helicity flux projected on right chirality ($\Pi_\Delta^{H1}$, $\Pi_\Delta^{H2}$) through helical wave decomposition [10] at the filter width $\Delta=48\eta$ are shown in Fig.6. Compared to the more flatter and coarser structure of $\Pi_\Delta^E$, the structure of $\Pi_\Delta^{H1R}$ is more tubelike, which is well reflected in the forward and backward cascade in Fig.6. Moreover, the spatial distributions of the backward cascade of $\Pi_\Delta^E$ and $\Pi_\Delta^{H1R}$ are relatively different, and the distribution of backward energy flux is more concentrated and occupies a smaller region. In contrast to $\Pi_\Delta^{H1R}$, a more plump geometry of $\Pi_\Delta^{H2R}$ can be seen with the naked eyes, and its spatial distribution is more concentrated for forward as well as backward cascade. The spherical structure of $\Pi_\Delta^{H2R}$ corresponds to a more intermittent property, which is consistent with conclusion in Fig.4.

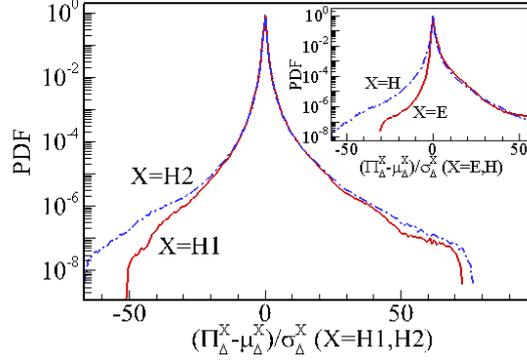

FIG.5. PDF of the first-channel helicity flux and the second-channel helicity flux with $\Delta=24\eta$. PDF of energy flux and total helicity flux with $\Delta=24\eta$ are shown in the inset.

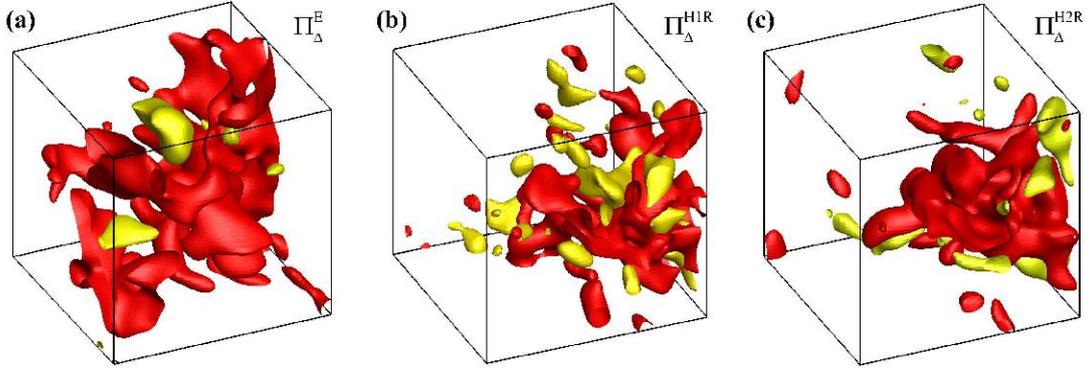

FIG.6. 3D views of the energy flux $\Pi_\Delta^E$ (a), the first-channel helicity flux projected on right chirality $\Pi_\Delta^{H1R}$ (b) and the second-channel helicity flux projected on right chirality $\Pi_\Delta^{H2R}$ (c) on the filter width $\Delta=48\eta$. They are rendered by $(X-m)/\sigma=2$ denoting forward cascade with red region, and $(X-m)/\sigma=-2$ denoting backward cascade with yellow region, where X represents variables $\Pi_\Delta^E, \Pi_\Delta^{H1R}$ and $\Pi_\Delta^{H2R}$, m and $\sigma$ are their mean and variance, respectively. Here, the involved regions contain only $256^3$ grid points.

Relative to triadic interactions of the same-chirality velocity [20], the dual-channel helicity cascade proposed in this letter provides a new perspective for the mechanism of hindered even inverse energy cascade. It is well known that the relative geometry of two tensors plays a basic role in the process of turbulent cascade, and it reflects the turbulent spatial structures to some extent. The relative geometries of energy flux, the first-channel helicity flux and the second-channel helicity flux can be defined as

$$\cos\theta^E = \frac{-\boldsymbol{\tau}:\tilde{\mathbf{S}}}{|\boldsymbol{\tau}||\tilde{\mathbf{S}}|}, \cos\theta^{H1} = \frac{-\boldsymbol{\tau}:\tilde{\mathbf{R}}}{|\boldsymbol{\tau}||\tilde{\mathbf{R}}|}, \cos\theta^{H2} = \frac{-\boldsymbol{\gamma}:\tilde{\boldsymbol{\Omega}}}{|\boldsymbol{\gamma}||\tilde{\boldsymbol{\Omega}}|} \qquad (6)$$

Their correlations are estimated numerically by joint probability distribution functions (JPDF) and conditional analysis on each other in Fig.7. The principal axis of JPDF of the correlation between energy flux and the first-channel helicity flux points toward to the upper right, which reflects the positive correlation of them. While the principal axis of JPDF of the correlation between energy flux and the second-channel helicity flux points toward to the upper left and lower left because of symmetric chirality, and it exhibits the negative correlation between energy flux and the second-channel helicity flux. This negative correlation means that energy cascade could be hindered or even inversed when the second-channel helicity flux is strong enough.

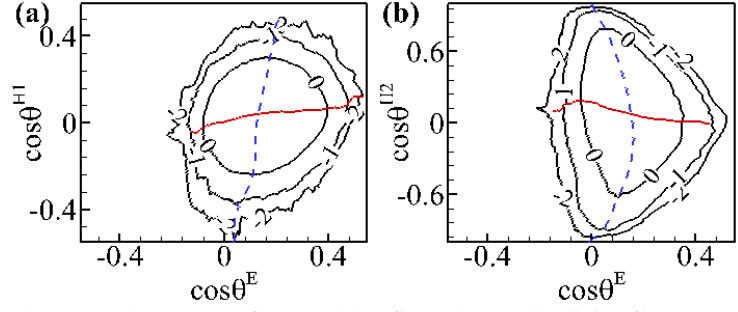

FIG.7. JPDF between the energy flux and the first-channel helicity flux (a) or the second-channel helicity flux (b). Red solid line are conditional averages of the abscissas conditioned on the ordinates and blue dashed lines are vice versa. $\Delta=96\eta$ is selected as the filter width.

In order to estimate the correlation between energy flux and helicity flux further, we take the local spatial average of energy flux conditioned on the first-channel and second-channel helicity flux with different filter widths in Fig.8. From the point of the second-channel helicity flux, the impact of helicity cascade on energy cascade changes from hindering to reversing with the decrease in characteristic length scales, which is obviously represented by large blue regions in Fig.8(b). On the contrary, the first-channel helicity flux always promotes the forward energy flux in all length scales, which is reflected by scarlet regions in Fig.8(a). Consequently, we could conclude that the first channel promotes the energy flux while the second channel hinders the forward energy cascade and even promotes the backward energy cascade.

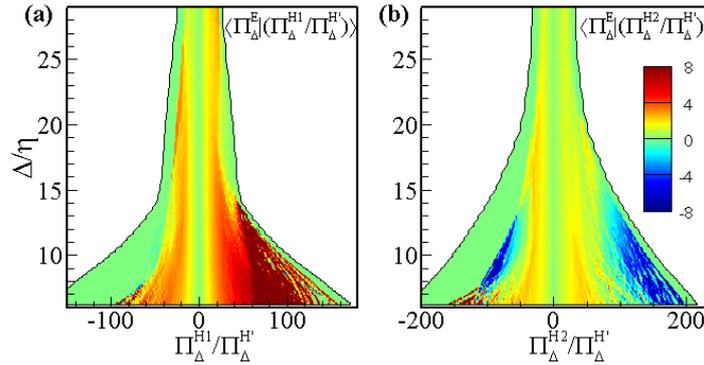

FIG.7. The local spatial average of energy flux conditioned on the first-channel helicity flux $\Pi_\Delta^{H1}$ (a) and the second-channel helicity flux $\Pi_\Delta^{H2}$ (b). The longitudinal axis represents different filter width.

This research reveals that there exists a dual-channel of helicity cascade from larger scales to smaller scales in turbulent flows, and these two channels correspond to vortex twisting and stretching processes, respectively. They behave differently depending on types of turbulent flows. In the analysis of helical homogeneous and isotropic turbulence, we find that they have different statistical properties, especially in high-order structure functions, probability distribution function and different spatial morphological characteristics. The newly proposed second channel of helicity cascade has a more obvious influence on the energy cascade, and it could be recognized as a new promoting mechanism for the backward energy cascade. The dual-channel helicity cascade might be used for suitably explaining some natural phenomena related to helicity, and should be considered for dealing with subgrid models based on helicity cascade in turbulent flows.